# Complete 2mm Spectral Line Survey (130-170 GHz) of Sgr B2N, Sgr B2OH, IRC +10 216, Orion (KL), Orion-S, W51M, and W3(IRS5)

A. J. Remijan (NRAO), D.P. Leigh (NRAO)
A. J. Markwick-Kemper (University of Manchester)
B. E. Turner (NRAO)

## *Abstract*

We report a complete 2mm spectral line survey (130-170 GHz) taken with the NRAO 12m Telescope between 1993 and 1995 toward the following sources: Sgr B2N, Sgr B2OH, IRC +10 216, Orion (KL), Orion-S, W51M, and W3(IRS5).

Until very recently, this project was entirely the work of B. E. Turner. He wrote the original proposal, given below without changes or updates, and did all of the observing. B.E. Turner has fallen seriously ill and can no longer continue to work on the analysis of these data. The notes that follow the proposal give further information about the project and important information for users of these data.

The data are distributed using the Spectral Line Search Engine (SLiSE) developed by A. J. Remijan and M. J. Remijan. SLiSE is a data display tool that will contain all the fully reduced and calibrated archived data taken as part of this 2mm survey. SLiSE is fast, easy to use, and contains the necessary functionality to display the data taken from spectral line searches. For example, SLiSE contains functions to overlay possible molecule identifications based on a current line catalog as well as overlaying H and He recombination lines. It is a Java-based applet, so it is platform independent and easily accessed online. The only caveat is that SLiSE was built using Java 1.5, so an update to the user's Java may be necessary.

We request users of these data to give B.E. Turner and this work the appropriate citation and credit.

# A Proposal for a 2mm Wavelength Spectral Survey at the 12m Telescope

B.E. Turner, NRAO
T. Amano, L.W. Avery, and P.A. Feldman, Herzberg Institute

## *The Importance of Spectral Surveys*

Unbiased, high-resolution, wide-band spectral surveys of molecular lines have been a major source of new information for interstellar chemistry and astrophysics. The published surveys to date include:

| | |
|---|---|
| Orion and IRC10216 | 72-90 GHz Johansson et al. 1984 |
| Orion | 215-263 GHz Sutton et al. 1985; Blake et al. 1987 |
| SgrB2 | 72-115 GHz Cummings et al. 1986 |
| SgrB2 and Orion | 335-360 GHz Jewell et al 1989 |

In addition, there are two unpublished surveys:

| | |
|---|---|
| Orion | 83-109 GHz Nobeyama |
| Orion and IRC10216 | 2mm window IRAM 30m |

The many lines observed in these surveys for each molecular species have allowed greatly improved estimates of abundances, or rotational temperatures, and hence of the physical conditions under which each molecular species occurs. Large numbers of unidentified lines have been found, leading to the identification of progressively more complex species. Other information available from the strengths of molecular lines bears on the densities and excitation conditions (both collisional and radiative) within molecular clouds. Further information is provided on gas kinetics, particularly in the regions such as Orion where line widths and central velocities have helped identify several distinct components of the gas, the so-called ambient cloud, hot core, and plateau components. Finally, preliminary surveying has begun at SEST, JCMT, and CSO at different positions within the giant molecular clouds such as SgrB2 and Orion, and is revealing striking differences in the chemistry over small distances.

The partial duplication of the existing 3mm surveys has proved very important for several reasons:

1. Much of the most interesting information is at the threshold of sensitivity of these surveys. Duplication has greatly increased the reliability of spectral lines near the sensitivity limits.

2. The number of U lines has depended strongly on the differing parameters of the various surveys. For example, the 3mm survey of Orion rank in order of increasing sensitivity to flux as NRAO/Onsala/Nobeyama but in order of increasing sensitivity to brightness as Nobeyama/Onsala/NRAO. The number of detected U lines follows the latter rank order rather than the former.

3. Surveys utilizing different beam sizes but similar sensitivity may be compared in terms of observed $T_R^*$ to give information on the source angular size. Thus, the Cummins et al. and Turner surveys of SgrB2 have revealed which molecular species are extended relative to the 60" NRAO beam, and which are not, and have allowed an actual estimate of source sizes, essential to determine abundances in the case of unresolved sources.

## *Some Important Specific Results from Previous Surveys*

### 1. Identification of New Species:

Several species were identified by isolating harmonically related sequences of lines in frequency. These include $HOCO^+$, $HCS^+$, $CH_3SH$, HNCS, $C_4H$, $C_3N$ from the Cummins et al. survey; and $CH_3NC$, $C_5H$, $C_6H$ and the remarkable metallic compounds NaCl, NaF, AlF, KCl from the IRAM 2mm survey; and finally several tentative identifications ($CH_2CH$-HCO, $CH_2CH$-NCH, $CH_3CH$=CHCN, $CH_2$=CHCN, $CH_2$=CH-$CH_2CN$) from the Turner survey, which require more accurate laboratory spectroscopy in order to confirm them. In addition, the first identified phosphorus compound, PN, was suggested by Sutton et al. on the basis of a line detected at 234 GHz in the CIT Orion Survey.

### 2. Fundamental Revision of Molecular Abundances:

Because of its high sensitivity to brightness temperature, the NRAO 3mm survey has detected many weak transitions of complex species (in the sense of small line strengths or small dipole components) not previously seen. When plotted on a rotation diagram, these transitions all lie systematically above LTE line, by large factors (10 to 100). Species involved are CH3OHCO, CH3CHO, EtOH, (CH3)2O, EtCN, VyCN. Either large optical depths or systematic pumping effects of the weak transitions are indicated. Attempts to simulate pumping effects have, not unexpectedly, failed. If large optical depths are involved, these complex species are typically 100 times more abundant than previously recognized, in both SgrB2 and Orion. Because such complex species are all believed to form via radiative association reactions, and because current laboratory and theoretical information suggests that the rates of many of these reactions may be smaller than presently adopted, any such upward revision of complex molecular abundances will have major implications for the astrochemistry of all large species. For example, it may suggest that such species are formed on grains rather than via gas phase reactions.

### 3. The Integrated Flux of Spectral Lines

The high density of lines present in rich sources such as Orion (KL) has been recognized as contributing significantly to the net continuum of such sources at millimeter wavelengths. Sutton finds that fully 40 percent of the total Orion (KL) flux arises from lines in the 1.3mm window, an amount that has major impact upon understanding the role of and of free-free emission in such a source.

### 4. Chemical Differences within Star Forming Regions

The CIT survey of the Orion core region was particularly successful in revealing clear contrasts in abundances among the hot core, plateau, compact ridge, and ambient cloud components. This was well accomplished at the higher frequencies of the CIT survey because the high excitation conditions in the hotter core and the plateau produce stronger line brightnesses which overcome the higher beam dilution of these spatially smaller components. In this way, Blake et al. (1987) have suggested, for example, that the energetic outflow from the hot core/IRc system impinges on the compact ridge, producing copious $H_2O$, with via radiative association reactions goes on to produce the large observed abundances in his region of species such as $CH_3OHCO$, $(CH_3)_2O$, $CH_3CHO$, EtOH, etc. Surveys of a wider region in Orion have more recently been undertaken (but are as yet unpublished). Comparative surveys in the 300-360 GHz region at CSO reveal that the two prominent CS sources north and south of IRc2, earlier studied by Mundy et al. (1988), are quite lacking of a rich chemistry, and appear to exhibit only a few simple, abundant species. The peculiar object known as Ori(1,3), which is particularly strong in certain ions such as $N_2H+$ and radicals such as $C_2H$, has not yet been surveyed.

Similarly, unpublished work at the JCMT, SEST, and CSO is revealing striking differences between the chemical nature of the SgrB2(OH) and SgrB2(N) components, which are separated by 80 arcsec. It has been known for some time that the SgrB2(N) exhibits much higher excitation than SgrB2(OH), but the question of its relative chemical richness is still open. The intermediate position SgrB2(M) is also now being surveyed in the 330-360 GHz region, and appears chemically richer that SgrB2(OH) although the lines are generally weaker. A start has been made on W51M by the CIT group. Partial coverage in the 330-360 GHz region has been obtained, and W51M appears to resemble Orion divided by 4, at least in terms of intensities. Finally, some very preliminary work by the CIT group has been done on such disparate objects as IRAS16293 and CRL2688, but these objects appear quite empty.

## *The Importance of a 2mm Survey*

No published 2mm survey currently exists. The 2mm window (125 – 175 GHz) is roughly the same width (50GHz) as the well-studied 3mm window (45GHz). In high-excitation sources such as Orion the 2mm window will contain more detectable lines (and hence more information) than the 3mm window, because of larger line strengths on average, yet fewer lines than the 1.3mm window, which is so crowded that much of the information cannot be used because of severe difficulties in deconvolving DSB data and because of inherent line blending. In lower excitation sources such as SgrB2(OH), the line density will be comparable to that of the 3mm window, and higher than in the 1.3mm window where the source is beginning to run out of excitation. In IRC10216, little is known outside the 3mm window, but the nature of the excitation in this source suggests the 2mm window may be the most useful for two reasons: i) Lines originating in the inner regions of the CSE have high enough excitation for the 2mm region but are badly beam diluted at 3mm, less so at 2mm; ii) Lines originating in the outer envelope depend on IR excitation, the mechanism of which insures a low excitation temperature (~20K) and hence inadequate excitation temperature for the 1.3mm window. The 2mm window should therefore provide the most information for inner core and outer envelope taken together. For the W51M source, little is known. Partial (unpublished) surveying at 3mm at the NRAO 12m telescope suggests the W51M is intermediate between Orion (high excitation) and SgrB2(OH) (lower excitation), and therefore that the 2mm window may be optimum.

## *Selection of Sources, Positions, and Frequency Coverage*

From the forgoing, it is obvious that the astrochemistry most interesting regions are Orion, SgrB2, W51M, and IRC10216, and we shall confine our proposal to these regions, which occupy complementary positions in the sky.

At present, the overwhelming body of astrochemical knowledge is based on only two sources: Orion(core) and SgrB2(OH). How typical are these sources? As mentioned, different positions within each of the sources reveal quite different chemical abundances. Even greater differences appear to exist between entirely different star forming regions such as SgrB2 and Orion. W51M appears to be quite different again. Finally, IRC10216 serves to define the differences between circumstellar and interstellar chemistry.

Time restrictions prevent coverage of the entire 2mm window for more than one position in each of the four proposed sources. We consider the importance of more than one position in Orion and SgrB2 to outweigh the advantages of full coverage of the 2mm window for at least two reasons: i) More than one position serves the same purpose as an additional source, given the steep gradient in the chemistry now recognized within such regions as Orion and SgrB2. It also gives additional information on just how deep such chemical (and excitation) gradients can be, of obvious importance in chemical modeling; ii) Part of the 2mm window (~60 percent for the region between 124 and 144 GHz) has already been observed by Cummins at al. in SgrB2, although the rather large beam (1.6 arcmin) at least partially smears the SgrB2(OH) and SgrB2(N) positions. In addition, there exists the unpublished IRAM survey (total frequency coverage unknown to us) which covers only the single KL position in Orion.

Our first priority is to survey two positions in each of Orion and SgrB2. Based on the results of the several surveys in progress, the desired positions in Orion are “KL” and (1,3). This clump at position (1,3) is the most massive of the four lying along the north-south Orion ridge and including KL, and it also has the lowest kinetic temperature (20-30K) as well as very high density (~$10^6$ $cm^{-3}$) and peculiar dust characteristics. From the small linewidths, low temperatures and high density, it appears to be a candidate for the birthplace of OB stars, that is to be an earlier evolutionary phase than Ori(KL). Every indication is that it produces a unique chemistry; it is particularly rich in molecular ions and free radicals, surprising though that they may be considering the high densities. The two positions in SgrB2 are “OH” and “N”. The OH position is much richer in lines at 3mm, but the N position is much richer at 1.3mm. Only part of the difference seems attributable to differing excitation conditions.

We have examined the Lovas catalog of spectral lines, as well as compilations of frequencies we have calculated for various other prospective interstellar species not yet identified. We conclude that the information content within the 2mm window favors the higher end, though not decisively. The statistics of the line distribution, as well as the realistic time constraints, indicate that a 35 GHz region, from 140 to 175 GHz can be covered for the two positions in Orion, and a smaller region, from ~150 to 175 GHz fro the two positions with SgrB2, as well as W51M and IRC10216.

In addition to the general wealth of information that will accrue from the proposed survey, the proposed region of the coverage will allow answers to the following questions:

1. Many more even weaker transitions will be detected from the complex molecular species which showed such transitions to be anomalously strong in the NRAO 3mm survey. The proposed spectral resolution (1.3 km/s), as well as the high special resolution, will help decide whether these overly-strong "weak" transitions are in fact (weak) masers, or whether their surprising strength is instead due to large optical depth, which would require and upward revision of the abundances of these complex molecular species by a factor of ~100.

2. Several prospective harmonic sequences currently recognized in the 3mm survey but not verifiable in the 1.3mm window because of line crowding, may be verified in the 2mm window. This should lead to identification of new molecular species.

3. The higher spatial and frequency resolution for the proposed survey relative to that of previous surveys will help to clarify the physical environment in which many molecular species arise and hence to understand their chemistry better. The multi-component Orion KL core region is the case in point. Interferometry is ultimately the answer to these questions, but interferometry is not suited to weak, extended emission, and detailed study of the large numbers of transitions necessary for an overall perspective is not feasible with current instruments.

4. The 2mm survey will be more sensitive to brightness than all existing 3mm surveys, yet not suffer the line crowding of the 1.3mm surveys. Thus it has the potential to reveal much new information which is fundamentally absent from other surveys. In Orion, for example, we expect that many more U lines will be seen per frequency interval, because of the larger line strength at 2mm compared to 3mm, and because the effective brightness limit at 1.3mm is seriously limited by line crowding from know species. It is expected that the 2mm SIS receiver will be tunable SSB, greatly reducing the threat of line crowing even further.

## *Survey Parameters*

The combination the new hybrid spectrometer and the new 2mm SIS junction receiver at the NRAO 12m telescope provide a unique opportunity for the 2mm spectral survey. The SIS receiver will permit a survey of unprecedented sensitivity to brightness temperature. The spectrometer, with 1536 channels, will allow such a survey to be done quickly.

The 2mm SIS junction receiver will be dual channel, and will cover the range 130-175 GHz with a hoped-for receiver noise temperature of typically $\leq$ 75K (DBS), with a worse case value of 100K DBS. The receiver is expected to be tunable SSB.

The hybrid spectrometer will be used with a bandwidth of 600 MHz, and provide 768 channels for each of the two receivers. The spectral resolution is thus 0.781 MHz/channel or 1.3 km/s at 150 GHz.

In 1 hr total integration (30 min on-source), a sensitivity of 66 mK (5-sigma) in $T_R^*$ units will be achieved, assuming $T_{sys} = (2T_R + T_{atm})\exp(\tau)/\eta_1\eta_{fss} \simeq 500K$ under good weather conditions. By comparison, the NRAO 3mm survey, the most sensitive in $T_R^*$ so far, achieved ~80mK (5-sigma) on average, although portions were up to 2 times less sensitive. The stronger lines expected at 2mm thus provide a great increase in S/N, hence the number of detected lines, over all previous surveys.

We propose to cover the 140-170 GHz region, a range of 35 GHz, in the following objects:

| | | |
|---|---|---|
| Orion (KL) | 05 32 47.0; -05 24 26 | 8hr/day |
| Orion (1,3) | 05 32 49.0; -05 21 20 | 8hr/day |
| IRC10216 | 09 45 14.8; +13 30 40 | 5hr/day |
| SgrB2 (OH) | 17 44 11.0; -28 22 30 | 5hr/day |
| SgrB2 (N) | 17 44 09.5; -28 21 15 | 5hr/day |
| W51M | 19 14 26.3; +14 24 43 | 4hr/day |

At 600 MHz per receiver setting, and 1 hr each, 14.6 days of Orion time is required to cover the two positions, with no overhead. A realistic overhead of 20 percent is needed for tuning, equipment and weather problems, and another 12 percent for telescope switching between on-source and reference positions. Thus 20 days is required to meet the goal for Orion. The time requirements for the two positions in SgrB2 become a little unrealistic (31 days) for the same frequency coverage, so for SgrB2 we propose instead to cover a smaller frequency range, namely 150-172 GHz, for each of the two positions. The same applies for IRC10216, and yet a smaller range for W51M (155-173 GHz). In choosing these parameters, two points should be noted:

1. Choice of a shorter integration time (e.g. 40min) and a larger frequency range becomes inefficient because of the overhead of tuning the receiver for each setting.

2. Double sideband operation requires the same number of receiver settings as SSB (since the LO must be shifted for each setting to deconvolve the SBs). Although the shifted setting might required less integration time, some setting may require more than one shift if the spectrum is unusually complicated. The overall time would be roughly the same, while the DSB mode certainly lacks the reliability of the SSB mode.

## *Other Considerations*

The proposed team of investigators includes the following experience relevant to understanding this survey:

1. Many U lines will be detected. We will rely on the intuition, experience, and laboratory skills of one of us (T.A.) to facilitate the identification of such lines, and to provide backup laboratory measurements where appropriate.

2. One of us (B.T.) has recently published the NRAO 3mm survey, and has developed extensive software to handle such surveys, both the graphical presentation of the data, and the analytic facilities or abundance and excitation determinations.

3. All four proposers are experienced in the astrophysics and astrochemistry, as well as in millimeter wave observing techniques. One of us (L.A.) is the Principle

Investigator of the project to survey IRC10216 in the 0.87mm window, using the JCMT.

4. Although one other (unpublished) 2mm survey exists of IRC10216 and (in part) of Orion, the present survey emphasizes several different sources and/or positions not covered in the previous survey. Where it duplicated the previous survey (IRC10216), we have argued the merits of duplication of such surveys when the entail different telescope parameters.

## *Summary*

The proposed 2mm spectral survey requires:

1. The 2mm SIS dual channel receiver, tunable over 140-175 GHz.

2. The hybrid spectrometer in 2*768 channel mode, 600 MHz bandwidth.

3. Twenty days, all LST, The atmospheric transmission being comparable at 3mm and 2mm, we assume this proposal could be carried out in the 3mm periods of the observing season (fall and late spring). The requested time could be granted in two blocks of ten days each if convenient, one block in the fall and the other in the late spring.

# Notes: Updates to the Proposal

## *Other Available Surveys*

While many spectral surveys of the sources explored by this work have been published since the original proposal was submitted, a spectral analysis in the 2mm range still remains incomplete. This survey presents data that will not only provide a more comprehensive understanding of the molecular complexity within these sources, but also will be a valuable comparison resource to published surveys in the regions of their frequency overlap. First, Lee (2001, 2002), and Ziurys and McGonagle (1993) all conducted relatively large frequency range surveys of Orion within the 2mm window using the Taeduk Radio Astronomy Observatory 14 m telescope and the FCRAO 14 m telescope, respectively. Examination of these data in the same frequency ranges could provide confirmation of the line identifications in these surveys and possibly provide more evidence for the presence of some less intense and as yet unidentified lines. In addition, the spectral analysis would yield a column density and rotational temperature for each identified molecule, which will be of great interest to compare with previously published parameters. If the values were found to vary significantly between surveys, the validity and accuracy of the rotational diagram analysis method in general might be questioned. Second, Cernicharo (2000) surveyed a region of IRC+10216 that completely overlaps the current survey. The difference in beam size for the two surveys could furnish information about molecular distribution and chemical properties of IRC+10216 upon spectral comparison. The investigation possibilities mentioned for the Orion comparison also apply.

An updated list of published spectral surveys to date (including array surveys) are:

| | |
|---|---|
| Orion and IRC10216 | 72-90 GHz Johansson et al. 1984 |
| Orion | 34-50, 83.5-84.5, 86-91 GHz Ohishi et at. 1986 |
| | 47, 87 GHz Madden et al. 1989 |
| | 70-115 GHz Turner 1991 |
| | 86 GHz Plambeck et al. 1982 |
| | 87-108 GHz Friedel et al. 2003 |
| | 98 GHz Murata et al 1992 |
| | 138-151 GHz Lee et al. 2001 |
| | 150-160 GHz Ziurys and McGonagle 1993 |
| | 160-165 GHz Lee et al. 2002 |
| | 172-256 GHz Dickens et al. 1997 |
| | 190-900GHz (with gaps) Serabyn et al. 1995 |
| | 225-262 GHz Liu, S.-Y. et al. 2002 |
| | 215-263 GHz Sutton et al. 1985; Blake et al. 1987 |
| | 325-360GHz Schilke et al. 1997 |
| | 334-343 GHz Sutton et al. 1995 |
| | 455-507 GHz White et al. 2003 |
| | 486-492 and 541-577 GHz Persson et al. 2007 |
| | 607-725GHz Schilke et al. 2001 |
| | 795-903GHz Comito et al. 2005 |
| | 795-903 GHz Comito et al. 2005 |
| IRC+10216 | 28-50 GHz Kawaguchi et al. 1995 |
| | 129-172 GHz Cernicharo et al. 2000<br>206-232 GHz Mancone et al. 2005<br>222-268 and 340-365 GHz Avery et al. 1992<br>330-358 GHz Groesbeck 1994<br>1523-6977 GHz Cernicharo 1996 |
| SgrB2 and Orion | 70-115 GHz Turner 1991<br>335-360 GHz Jewell et al 1989 |
| SgrB2 | 70-150 GHz Cummins, Linke & Thaddeus 1986<br>218-263 GHz Nummelin et al. 1998, 2000<br>330-355 GHz Sutton et al. 1991<br>1530-6380 GHz Polehampton et al. 2007 |

## *Initial Analysis of the B.E. Turner Survey Data*

As there is such an enormous quantity of data, time restrictions have limited our initial analysis to Orion(KL) and IRC+10216. With Orion(KL), column density and rotational temperatures were derived for the previously identified species $CH_3CHO$, $CH_3OCHO$, $CH_3OCH_3$, HNCO, $NH_2CHO$, $SO_2$, and SO serving as a comparison source to previous surveys. Work on the IRC+10216 data is less mature and is focused on a comparison with the Cernicharo et al. (2000) survey. In addition, we found various inconsistencies throughout the database that cannot be explained unless a full data reduction is performed on the raw data. These include a 100 MHz frequency shift to higher frequencies between 130 and 130.5 GHz for the following sources: Orion(KL), Sgr B2(OH), Sgr B2(N) and W51(M). For example, the well determined SiO transition at 130268.7 MHz is shifted in those data to ~130368 MHz. In the data we present using SLiSE (see below), we have tried to correct these rest frequency errors so that each line between 130 and 130.5 GHz are at their correct rest frequency. Furthermore, there are several intensity problems in the Orion(KL) and Sgr B2(N) data that may be attributed to faulty flux or passband calibration. A more complete analysis on the other data searching for these intensity problems is necessary.

### Orion(KL)

The Orion(KL) spectra are very congested with lines leaving a low incidence of clean lines that can be attributed to a single species. However, with about 1200 lines over the entire frequency range, many species can be identified as undeniably present. Therefore, though many of the lines are likely due to molecular transitions of two or more different species, the shear volume of lines allows for the determination of some of the physical properties of these molecules. Column densities and rotational temperatures were derived in a manner that deviates from the traditional method, but that allows for more accurate values. Rather than assuming each line is due to any specific transition and using these labels to make a rotational diagram, the spectrum was manually searched to identify all the lines that a specific species might have contributed to. Utilizing the astronomical molecular catalogue, Splatologue, a transition-resolved collection of JPL, CDMS, and Lovas/NIST molecular transition, "clean" lines that seemed to be due only to the species under investigation were selected. Taking into account only these "clean" lines, a synthetic spectrum was derived by adjusting the parameters to make a best fit. Going through each of the lines individually to identify possible interference from other molecules is key to this method, but requires considerable time and effort.

The following column densities and rotational temperatures were found using the method described above.

| Species | Column Density ($cm^{-1}$) | Rotational Temperature (K) |
|---|---|---|
| CH3CHO | $5.0 \times 10^{13}$ | 40 |
| CH3OCH3 | $6.0 \times 10^{15}$ | 75 |
| HNCO | $7.9 \times 10^{14}$ | 200 |
| NH2CHO | $9.0 \times 10^{13}$ | 90 |
| SO2 | $3.2 \times 10^{15}$ | 70 |
| SO | $1.0 \times 10^{15}$ | 15 |

### IRC+10216

A vast majority of the spectral features detected in the Cernicharo et al. (2000) are also seen in the 12m data presented. However, we have not yet calculated the statistics on how many and what percentage of lines are detected uniquely between the 12m and 30m data. In addition, we are in the process of using the Beiging & Tafalla (1993) model to directly compare the spectra between the 12m and 30m data. This analysis is ongoing.

## *Accessing the B.E. Turner Spectral Data*

The data can be found in electronic form at:

www.cv.nrao.edu/~Turner2mmLineSurvey

The data are distributed using the Spectral Line Search Engine (SLiSE) developed by A. J. Remijan and M. J. Remijan. SLiSE is a data display tool that will contain all the fully reduced and calibrated archived data taken as part of this 2mm survey. SLiSE is fast, easy to use and contains the necessary functionality to display the data taken from spectral line searches. For example, SLiSE contains functions to overlay possible molecule identifications based on a current line catalog as well as overlaying H and He recombination lines. It is a java based applet, so it is platform independent and is easily accessed online. The only caveat is that SLiSE was built using Java 1.5. So an update to the user's java may be necessary. In addition, SLiSE will work on a Mac running Safari, not Netscape. Finally, the online version of SLiSE does have the capability to upload your own dataset to use the functionality that is offered for line identification.

Systematic errors in line frequencies and relative intensities are sure to exist based on the amount of time over which these data were taken. However, we do not have access to all of B.E. Turner's calibration and data reduction routines. We present the calibrated data using the SLiSE tool.

If you are interested in obtaining the raw data from the 12m during this time, we are working on making the fits data available at that same web location through the NRAO single dish data archive. This is currently a work in progress and subsequently may not be available until Fall, 2008. Any data requests then should be made to A. J. Remijan directly at: aremijan@nrao.edu. He will provide the publicly available raw data to a user.

## *B.E. Turner's Survey Summary: (survey.summary)*

In the following, we list the observation date, center frequency of the observation, the source observed and the appropriate scan numbers associated with each source. B.E. Turner used the following shorthand when writing out the sources:

IRC – IRC +10 216
SN – Sgr B2N
SOH – Sgr B2 OH
OS – Orion S
OKL – Orion KL
W51 – W51 Main
W3 – W3 IRS 5

The notes below have the following format:
Date
Frequency Source(scan#)

Observation summaries:

Jan/93
139.00 IRC(6-8) W51(9-17) SOH(18-26) SN(27-35) W3(4)(84-88) OS(113-120) OKL(121-128),IRC(167,179)
139.55 SN(36-44) SOH(45-53) W51(54-62) W3(4)(89-94) OKL(95-103) OS(104-111) IRC(158,166)
140.10 W51(63,72) OKL(129,137) OS(138,146) IRC(149,157)
140.65 W51 (73-81) IRC(180,189)
**all W51 no good-position error

Feb/93
140.10 SN(173,179) SOH(180,182
140.65 OS(156,164) OKL(165,168) SN(595,603) SOH(604,613)
141.20 W3(4)(129,137) OKL(138,146) OS(147,155)* OS(511,514) W51(568,576) SOH(577,585) SN(586,594)
141.75 W51(614,620)*
**all W51 no good-position error

Oct/93
139.00 W3(5)(903,911)
139.55 W3(5)(894,902) W51(1046,1054)
140.10 W3(5)(885,893) W51(1037,1045)
141.20 W3(5)(876,884)
140.65 W51(858,866) W3(5)(867,875)
141.20 IRC(5,13) W51(849,857)
141.75 IRC(14,22) W3(5)(127,135) OKL(136,144) OS(145,153) W51(233-236)** SOH(237,246) SN(247,255) W51(840,848)
142.30 IRC(23,31) OS(154,162) OKL(163,171) SN(256,264) SOH(265,275) W51(276,284)** W51(831,839)
142.85 IRC(35,43) W51(48,50)** SOH(51,59) SOH(69,71) SN(60,68) W51(109,116)** W3(5)(117,126) OKL(172,180) OS(181,189) W51(477,485)

143.40 SOH(72,81) SOH(94,96) SN(82,93) W51(100,108)** IRC(190,198)
W3(5)(303,311) OKL(312,320) OS(321,329) W51(468,476)
143.95 IRC(200,208) W51(285,293)** OS(330,338) OKL(339,347) SOH(436,444)
SN(445,458) W51(459,467)
144.50 IRC(209,217) W51(294,302)** OKL(348,356) OS(357,365) SOH(418,426)
SN(427,435) W51(651,659)
145.05 IRC(218,226) OS(366,373) OS(497,499) W51(415,417)** W3(5)(486,496)
OKL(500,508) W51(595,603) SOH(604,613) SN(614,622)
145.60 IRC(227,232) IRC(375,378) OKL(509,517) OS(518,526) SN(623,631)
SOH(632,641) W51(642,650)
146.15 IRC(379,387) OS(527,535) OKL(536,544) W51(661,669) W3(5)(670,676)
SN(813,821) SOH(822,830)
146.70 IRC(388,396) OKL(545,553) W3(5)(677,681) OS(682,690) W51(786,794)
SOH(795,803) SN(804,812)
147.25 IRC(397,405) OS(691,699) W51(977,985) SN(986,994) SOH(995,1003)
W3(5)(1055,1063) OKL(1064,1072)
147.80 IRC(406,414) IRC(555,557)* SOH(1004,1012) SN(1013,1027)
W51(1028,1036) OKL(1073,1081) OS(1082,1090
148.35 IRC(558,566)
148.90 IRC(567,575)
149.45 IRC(576,584)
150.00 IRC(585,594)
* not confirm line in Rx 1 of (406,414)

Jan/94

147.25 W3(391,399)
147.80 W3(403,411)
148.35 W51(11,19) W3(20,28) OKL(29,37) OS(38,46)
148.90 W3(48,56) OKL(57,65) OS(66,74)
149.45 W3(75,83) OKL(84,92) OS(216,224)
150.00 OKL(226,234) OS(235,243) W3(244,252)
150.55 IRC(93,101) W3(253,261) OKL(262,270) OS(271,279)
151.10 IRC(102,110) W3(412,420) OKL(421,429) OS(430,438)
151.65 IRC(111,119) OS(439,447) OKL(448,456)
152.20 IRC(280,288) OKL(457,465) OS(466,469)
152.75 IRC(292,300)
153.30 IRC(301,309)
153.85 IRC(470,478)
154.40 IRC(489,497)

Mar/94

148.35 SOH(58,66) SN(67,75) SOH(169,177)
148.90 W51(33,41) SN(42,50) SOH(51,57)** SOH(178,186)
149.45 W51(76,80)* W51(84,90) W51(161,168) SOH(187,194) SN(195,203)
150.00 W51(204,212) SOH(372,378) SN(379,387)
150.55 W51(213,221) SN(358,364) SOH(365,371)
151.10 W3(91,99) OKL(101,109) OS(110,118) W51(222,230)*** W51(343,348)
SN(529,537) SOH(538,546)
151.65 OKL(119,127) W51(231,233) W3(234,242) OS(243,251) W51(388,393)
SOH(547,555) SN(556,564)

152.20 OS(252,260) OKL(261,269) W3(270,278) W51(394,402)
152.75 W3(279,287) OS(288,296) OKL(297,305) W51(403,411)
153.30 W3(412,420) OKL(421,429) OS(430,438) W51(522,528)
153.85 IRC(2,10) OS(439,447) OKL(448,456) W3(457,465) W51(565,573)
154.40 IRC(13,21) OS(468,476) OKL(477,485) W51(574,582)
154.95 IRC(22,30) W51(583,591) W3(592,600) OKL(603,611) OS(612,620)
155.50 IRC(129,137)
156.05 IRC(138,146)
156.60 IRC(147,155)
157.15 IRC(158,160) IRC(308,313)
157.70 IRC(487,495)
158.25 IRC(325,333)
158.80 IRC(334,342)
159.35 IRC(496,504)
159.90 IRC(505,513)
160.45 IRC(516,521)
* data all bad (HS)
** data all likely bad
*** Rx 2 probably mistuned

Apr/94

154.40 W3(97;99,107)
155.50 W3(108,116) OKL(117,125)
161.00 IRC(126,134)
161.55 IRC(135,136)* IRC(137,144)
162.10 IRC(145,152)
162.65 IRC(153,161)
*Rx 2 mistuned (to DSB)

May/94

135.15 IRC(861,868)
135.70 IRC(852,860)
136.25 IRC(841,849)
136.80 IRC(832,840)
137.35 IRC(675,682)
137.90 IRC(665,672)
138.45 IRC(657,664)
152.20 SOH(169,176) SN(177,185)
152.75 SN(6,13)* SN(154,159) SOH(160,168)
153.30 SN(186,193) SN(346,348) SOH(337,345)
153.85 SN(349,357) SOH(358,366)
154.40 SOH(367,374) SN(521,529)
154.95 SN(530,538) SOH(539,547)
155.50 W51(14,16) OS(50,58) W51(147,153) SOH(548,556) SN(557) SN(692,700)
156.05 W51(17,27) W3(41,49) OS(59,67) OKL(68,76) SN(701,709) SOH(710,718)
156.60 W51(28,36) OKL(77,85) OS(86,94) W3(95,103) SOH(719,727) SN(728,732) SN(877,881)
157.15 OKL(104,115) W51(194,202) W3(223,231) OS(232,240) SN(882,890) SOH(891,899)

157.70 W51(203,211) OS(241,249) OKL(250,258) W3(259,267) SOH(900,908) SN(909,917)
158.25 W51(212,220) W3(268,275)** OS(276,281) W3(283,289) OS(438,446) OKL(447,455)
158.80 W51(328,336) W3(411,419) OKL(420,428) OS(429,437)
159.35 W51(375,383) OS(456,464) OKL(465,473) W3(585,593)
159.90 W51(384,392) W3(W3(594,602) OKL(603,611) OS(612,620)
160.45 W51(393,401) W3(402,410) OS(621,629) OKL(630,638)
161.00 W51(512,520) OKL(639,647) OS(648,656) W3(760,768)
161.55 W51(558,566) W3(769,777) OS(778,786) OKL(787,795)
162.10 W51(567,575) OKL(796,804) OS(805,813) W3(814,822)
162.65 W51(576,584) OKL(823,831)
163.20 IRC(116,124) W51(683,691)
163.75 IRC(128,136) W51(733,741)
164.30 IRC(137,146) W51(742,750)
164.85 IRC(290,298) W51(751,759)
165.40 IRC(299,307) W51(869,876)
165.59 W51(918,926)***
165.95 IRC(310,318)
166.50 IRC(319,327) W51(927,935)
167.05 IRC(474,482) W51(936,942)
167.60 IRC(483,491)
168.15 IRC(494,502)
168.70 IRC(503,511)
* scans 6-10 lost on HS
**pretty bad due to thick clouds
***freq. error; should be 165.95

Oct/94

131.85 IRC(383,391)*
132.40 IRC(374,382)
132.95 IRC(365,373)
133.50 IRC(198,206)
134.05 IRC(189,197)
134.60 IRC(180,188)
136.25 W51(1009,1014)
136.80 W51(1003,1008) W3(1088,1090) OKL(1091,1098)
137.35 W51(952,958) OS(1067,1073) OKL(1074,1080) W3(1081,1087)
137.90 W51(818,826) W3(1046,1052) OKL(1053,1059) OS(1060,1066)
138.45 W51(768,773) W51(814,817) OS(899,907) OKL(908,916) W3(1015,1020) W3(1039,1045)
142.30 W3(827,835)
143.95 W3(836,845)
144.50 W3(1025,1031)
145.60 W3(1032,1038)
158.25 SN(53,61)** SOH(62,70)** SN(995,997) SOH(998,1002)
158.80 SOH(32,43) SN(44,52)
159.35 W51(207,212) SN(213,222) SOH(223,231)
159.90 SOH(232,240) SN(241,249)
160.45 IRC(25,29) SN(397,405) SOH(406,414)

161.00  SOH(415,423) SN(424,433)
161.55  W51(580,582) SN(583,592) SOH(593,601)
162.10  SOH(602,612) SN(613,622)
162.65  W3(98,106) OS(107,115) SN(774,783) SOH(784,793)
163.20  OS(116,124) OKL(125,133) W3(134,142) SOH(794,802) SN(803,813)
163,75  W3(143,151) OS(152,160) OKL(161,169) SN(959,967) SOH(968,976)
164.30  W3(286,294) OKL(295,303) OS(304,312) SOH(977,985) SN(986,994)
164.85  W3(313,321) OS(322,330) OKL(331,339)
165.40  OKL(340,348) OS(349,355) W3(458,466)
165.95  W51(71,79) W3(467,475) OKL(476,484) OS(485,493)
166.50  W3(277,285) OS(494,502) OKL(503,511)
167.05  W3(2;68,276) OKL(512,520) OS(521,529)
167.60  W51(80,88) W3(89,97) OKL(530,536) OKL(715,720) OS(721,726) OS(892,898)
168.15  W51(250,258) W3(259,267) OKL(697,705) OS(706,714)
168.70  W51(392,396) W51(434,439) OS(679,687) OKL(688,696) W3(846,854)
169.25  IRC(6,14) W51(440,448) W3(652,660) OKL(661,669) OS(670,678)
169.80  IRC(15,23) W51(449,457) W3(643,651) OKL(855,863) OS(864,872)
170.35  IRC(171,179) W51(623,631) W3(632,642) OS(873,882) OKL(883,891)
170.90  IRC(356,364)
* tuned LSB
** weather bad

March/95 (regular assigned + NYA time)

131.85  IRC(209,217)
132.95  W51(481,485)
133.50  W51(377,383)
134.05  W51(370,376)
134.60  W51(319,321) W51(363,369)
135.15  W51(148,156)
135.70  W51(139,147) W3(384,390)
136.25  W3(163,171) OS(196,204) OKL(205,208) OKL(218,224)
136.80  W3(157,162) OS(187,195)
139.00  W51(130,138)
140.10  SOH(123,129)
140.65  OKL(181,186)
141.20  SOH(115,116) SOH(119,122) OS(172,180)
155.50  OS(225,233)
164.85  SOH(323,330) SN(331,338)
165.40  SN(339,346) SOH(347,354)
165.95  SOH(355,362) SN(428,436)
166.50  SOH(412,419) SN(420,427)
167.05  SN(396,403) SOH(404,411)
167.60  SN(439,450)** SOH(451,460)**
168.15  SOH(461,470)** SN(471,480)**
**weather bad

## *B.E. Turner's Survey Notes: (survey.notes)*

The following is a list of comments made by B.E. Turner to detail his thoughts while working on the project. These have been left in their original form and have not been updated. Perhaps these remarks will be of use to the reviewer of this survey. In addition, B.E. Turner was able to make some rough line identifications for Orion-KL and W51M which could be made accessible if necessary. Contact Anthony Remijan at aremijan@nrao.edu for more information about the line identifications.

## Notes for 2mm Survey, Compiled While Reducing Spectra

0) At the extremely low levels reached by this survey, spectra in objects like Ori(KL) become complex. Features recorded as lines depend on more than just intensity. Characteristic shapes are a criterion.
1) There are a class of features in Ori(KL) all of whose members have a highly characteristic profile, skewed. The paradigm of this class is the lines of $SO_2$. Examine all such lines (many are weak): are all of them $SO_2$? Can we say all species with such lines share a high-velocity component? The profile is skewed toward low freq (red) and sharp at high freq (blue). One possible fit to an $SO_2$ line (~135700) requires 4 gauss components. We assume the skewness is really all one species and do not define separate lines in our measurements. Thus the linewidth is listed as wide. Some lines (e.g. 142575 in Ori(KL) skew in the opposite direction.
2) In comparing lines seen at the high end of one spectrum and at the low end of the next spectrum, one should take the average of the two determinations.
3) In choosing line parameters in difficult cases, we compare "similar" objects such as Ori(KL) and W51M.
4) Particularly difficult are block-like profiles, almost squares; if W51M resolves them into two features, then we typically arbitrarily do the same in Ori. If there is a hint of 2 features, even though the detail is at the 1-sigma level, we usually decompose into >1 feature. Ultimately, an identification of the species involved should serve to resolve ambiguities.
5) We repair baselines and residual ramp effects while measuring lines. The repaired spectra are written over the originals, so the corrections will be in the WEB-stored data. It will not be corrected on the hardcopy pages. Question: are storage dates accessible in NSAVE files?
6) The MAC data is calibrated too low wrt filter bands, and the Hybrid Spectrometer (HYSPEC) data probably too high. We should compare several spectra taken with each to check. Unfortunately, the Millimeter AutoCorrelator (MAC) data was only taken to extend the low-freq end of the survey, or, on occasion, to improve low-quality HYSPEC data. The overlap is minimal.
7) Possibly discuss de-ramping, small-scale tinkering with baseline after the fact (while measuring lines).
8) Review the HYSPEC parameters (Hanning, 2-bit, 3-level, 0.809 factor, etc).
9) Some spectral features are deliberately not recorded if they are, e.g., dome-shaped and seem to be artifacts of the HYSPEC. They are not expunged from the spectra, however, but retained for the line identification program, where they may be useful.
10) Many lines are recorded that are below the 5-sigma noise level. Careful comparison of two very similar sources--W51M and Ori(KL) are used to decide on the reality of a line. It must appear in both sources, giving a redundancy of two receivers times two sources. Iteration was required. W51M was first analyzed, then Ori(KL), which has more lines, then W51M was completely redone, adding more lines that were marginal but which appeared in Ori(KL).

11) Ori(KL) vs. W51M: many lines are wider in W51M, some are about the same width, and some are much wider (and stronger) in Ori(KL). Keep track of these, and correlate them with species. (SO, $SO_2$ are examples of the third category).
12) Ori(KL) & W51M: W51M was done first, without comparison with Ori(KL). When Ori(KL) was next done, it was realized that the W51M recording of lines was highly conservative. W51M was then repeated in detail, adding lines that had counterparts in Ori(KL), and even some that didn't. Ramp errors were also corrected in the archival spectra, although these were rather few. In the comparison with Ori(KL), we spotted several serious ramp errors in Ori(KL). After the second reduction on Ori(KL) was finished, we went back and corrected ramp errors in Ori(KL). We did NOT intercompare systematically again with W51M, although a few spectra were compared. However, where ramp corrections were serious, we re-recorded lines, since intensities changed. The first Orion list should be considered somewhere between the two W51M lists in degree of conservatism. Because it was done after the first W51M list, it is more liberal (personal equation was shifting in that direction), but because few lines were subsequently added (no comparison with W51M) it is not as liberal as the second W51M list.

## Complete Survey Images:

Below is a compilation of the data available in the 2mm spectral line survey. The data presented below are a representation of the fully reduced, continuum-subtracted spectra performed by B. E. Turner. As stated above, the fully reduced, continuum subtracted spectra are available using SLiSE and the raw data on any passband can be obtained by contacting A. J. Remijan at aremijan@nrao.edu.

## W51 M (Turner) - 56 km/s (12m)

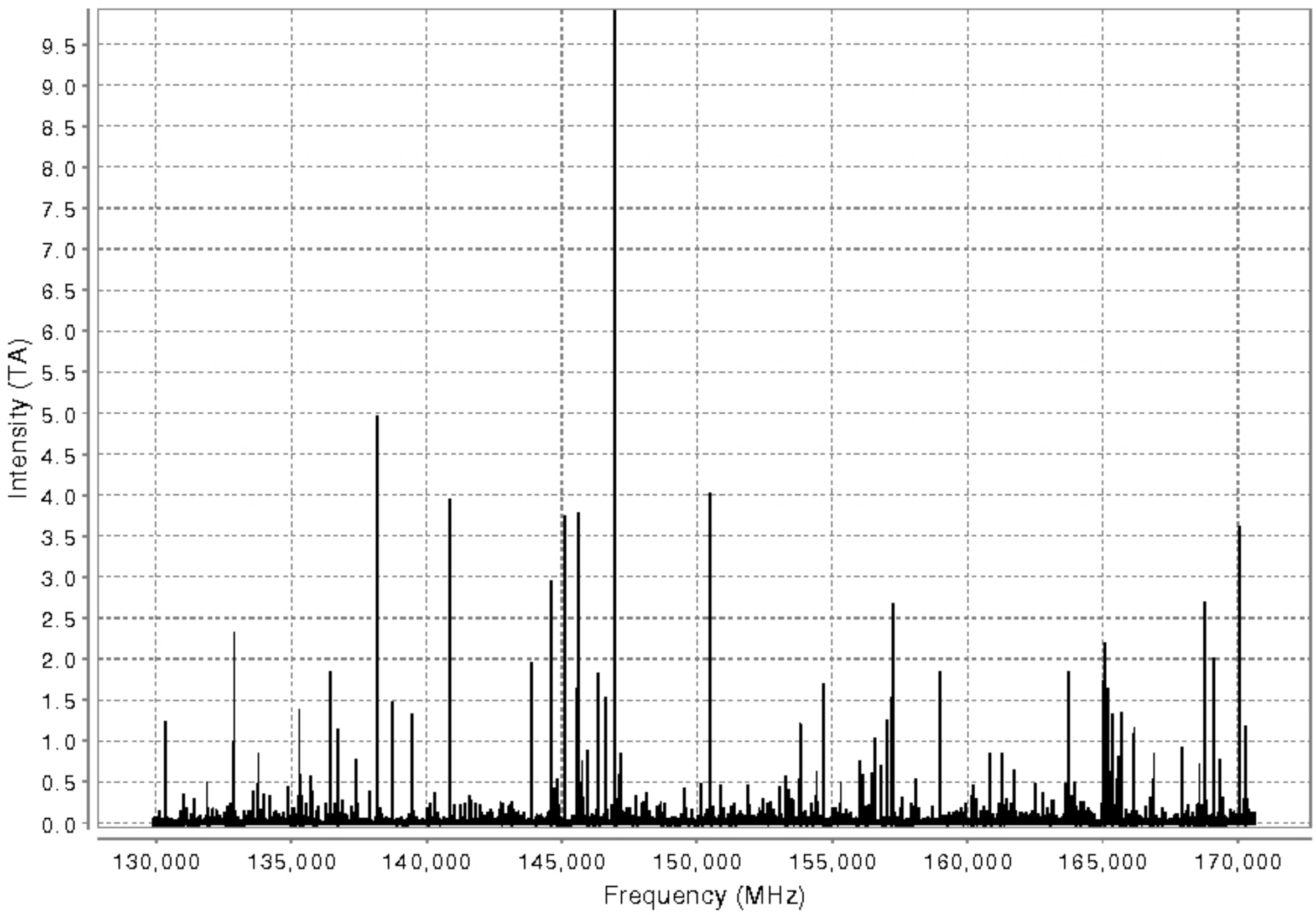

## IRC10216 (Turner) - -26 km/s (12m)

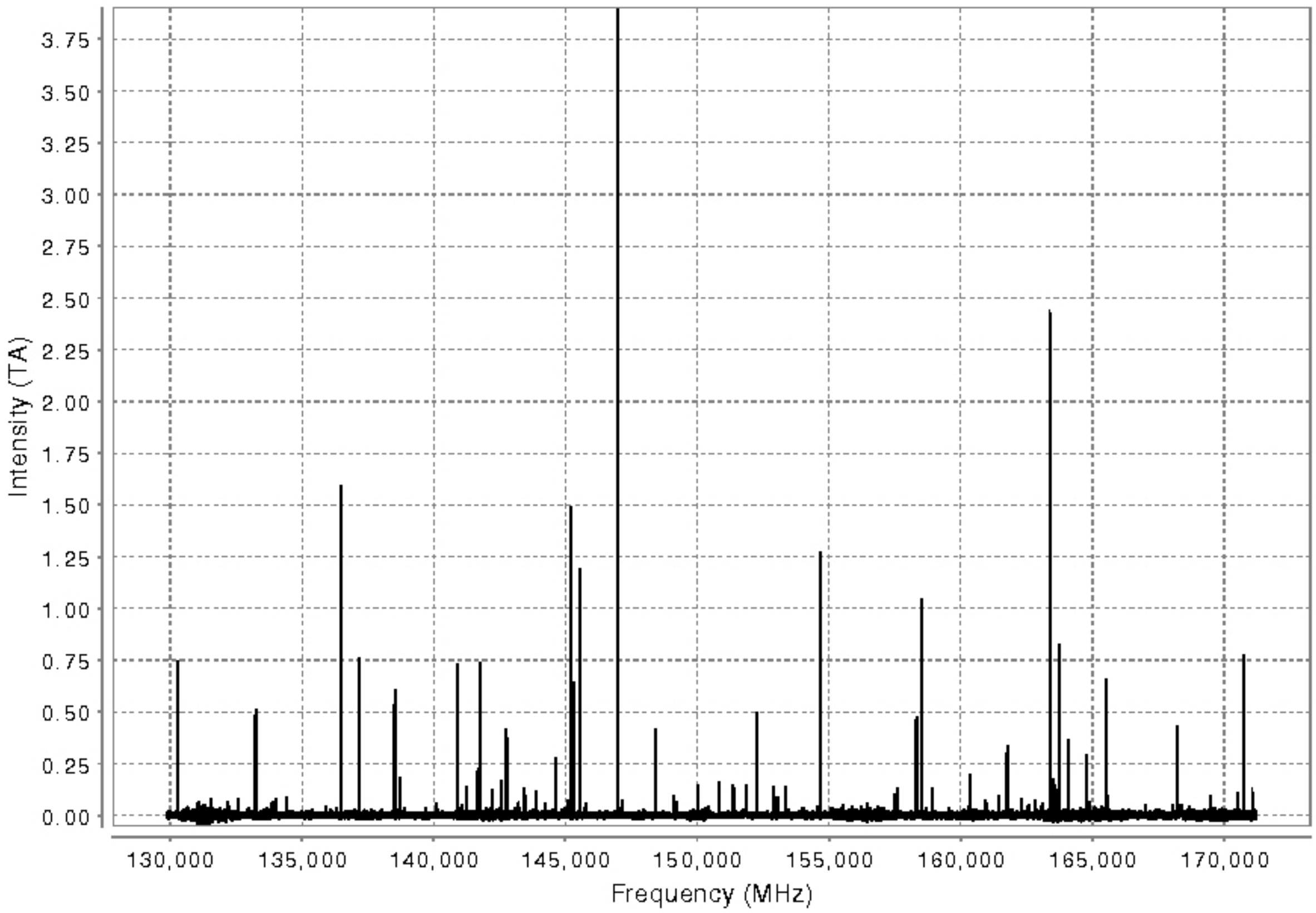

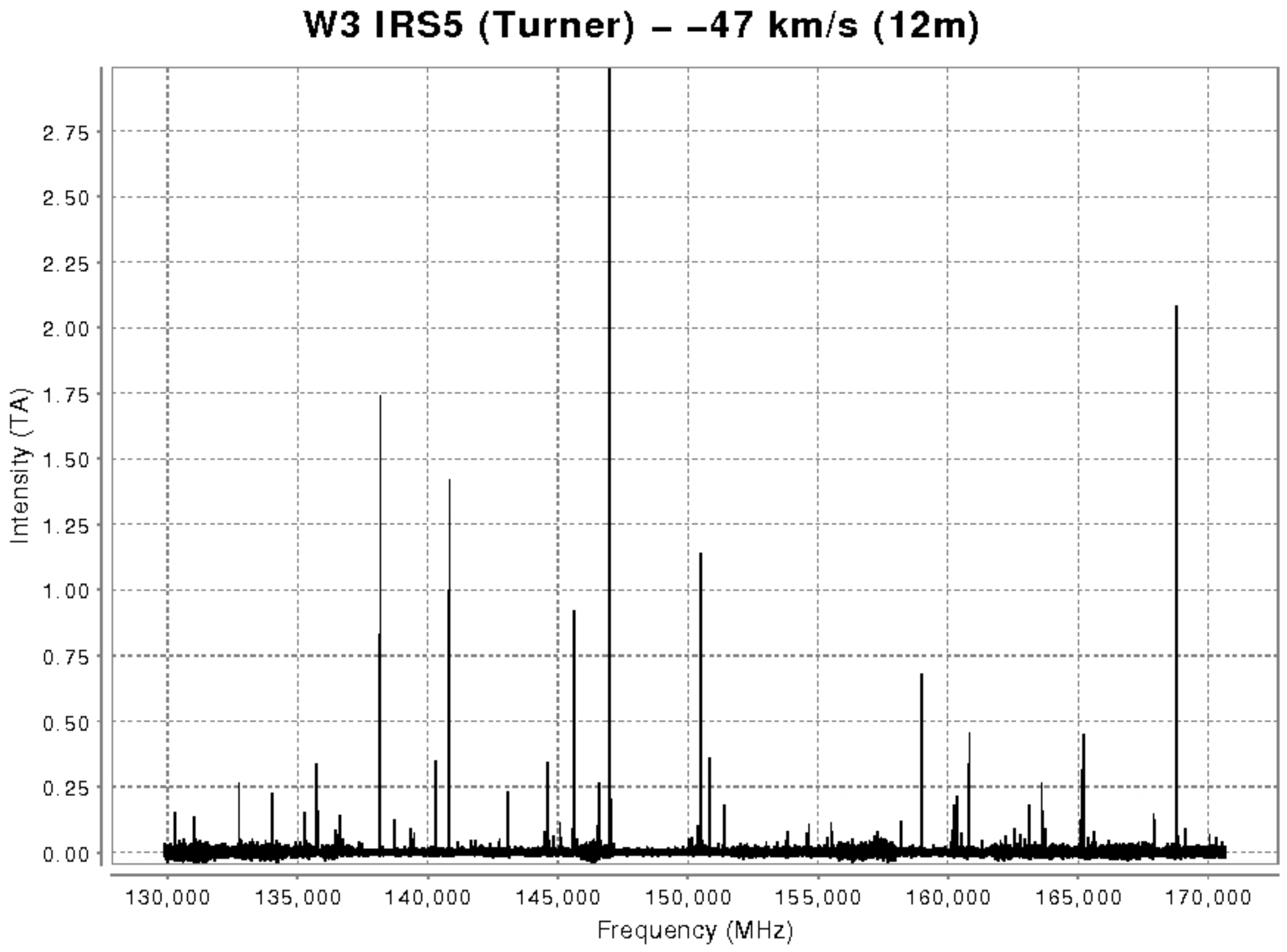

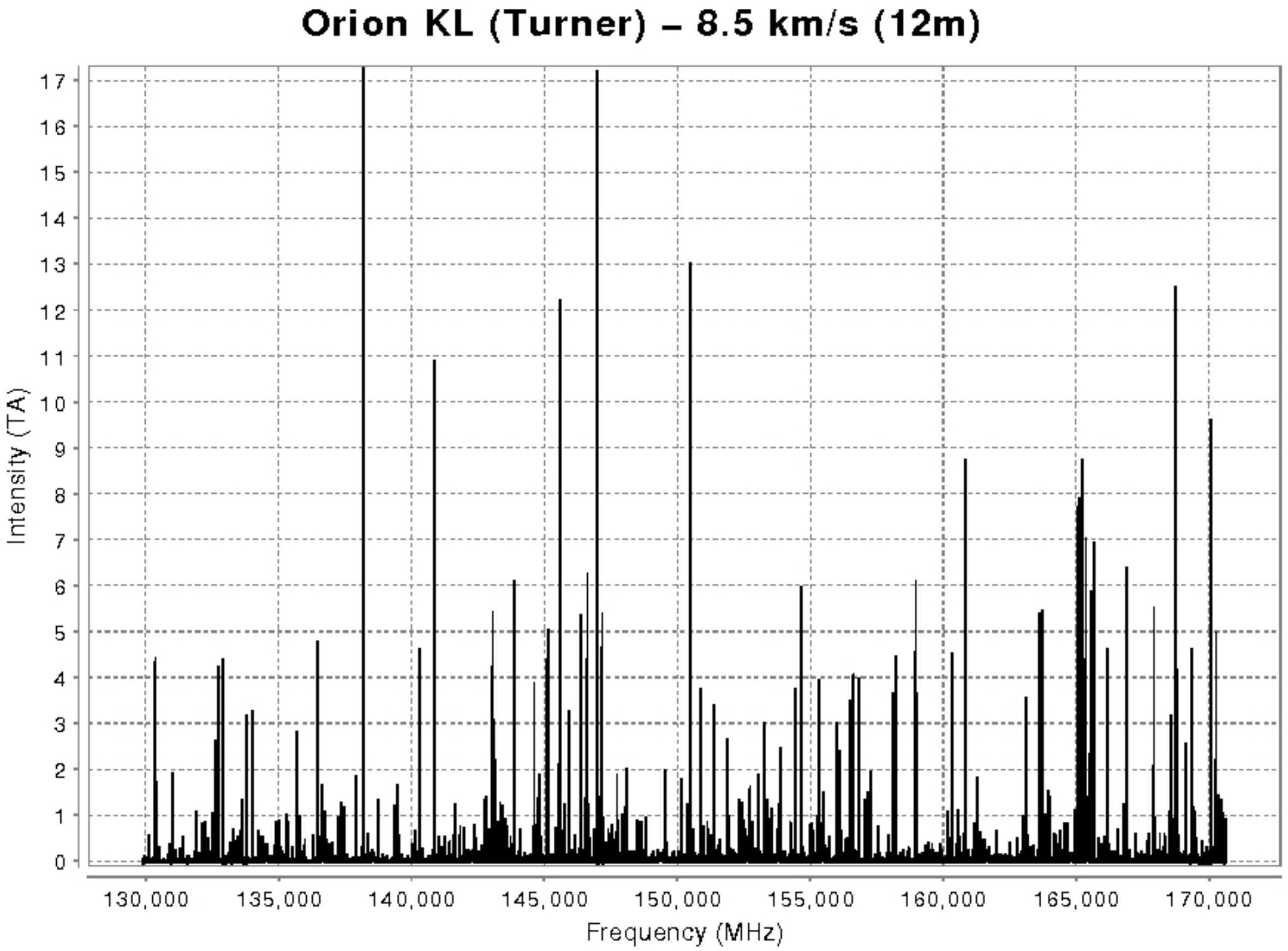

**SgrB2N (Turner) – 64 km/s (12m)**

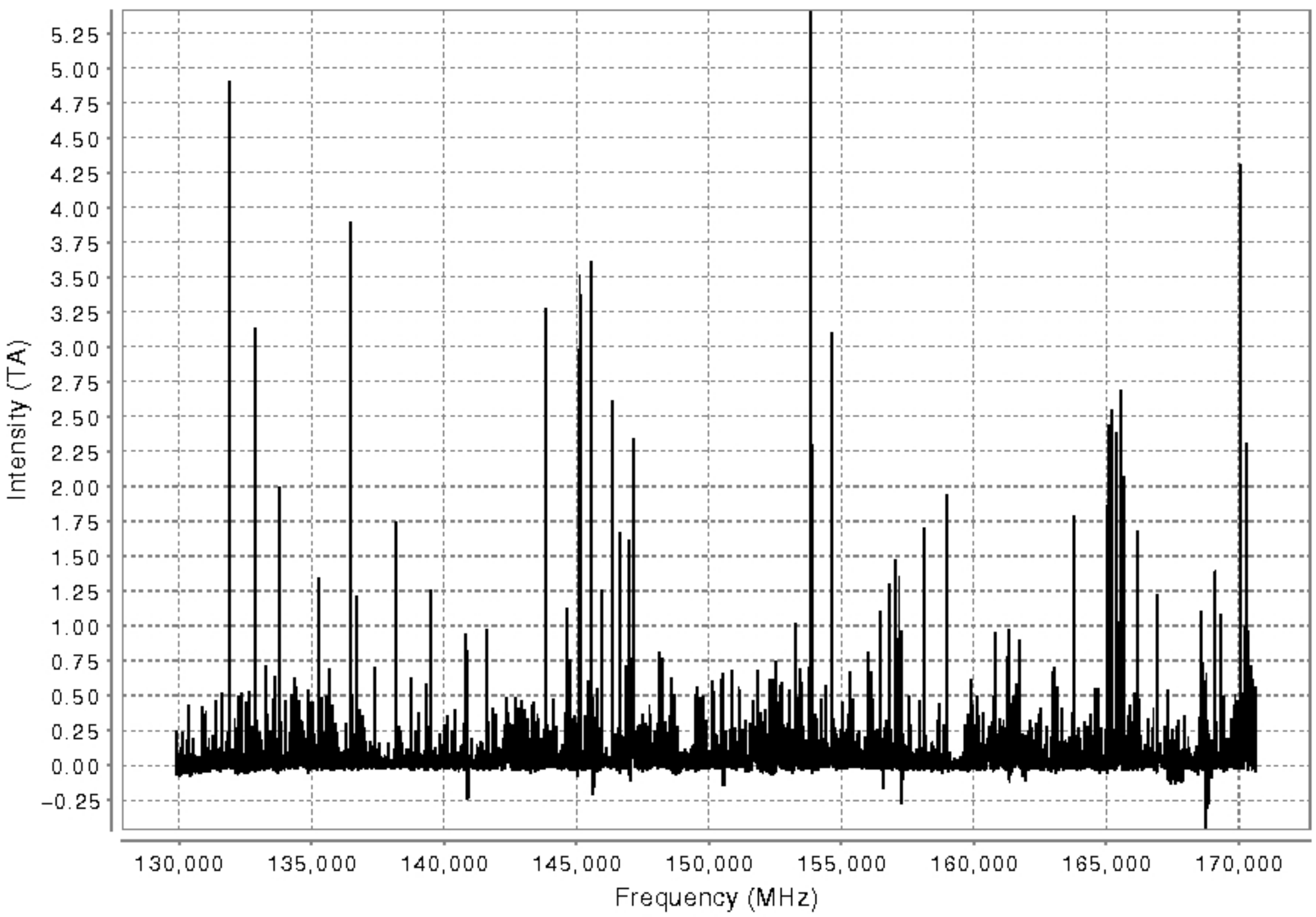

**SgrB2 OH (Turner) – 64 km/s (12m)**

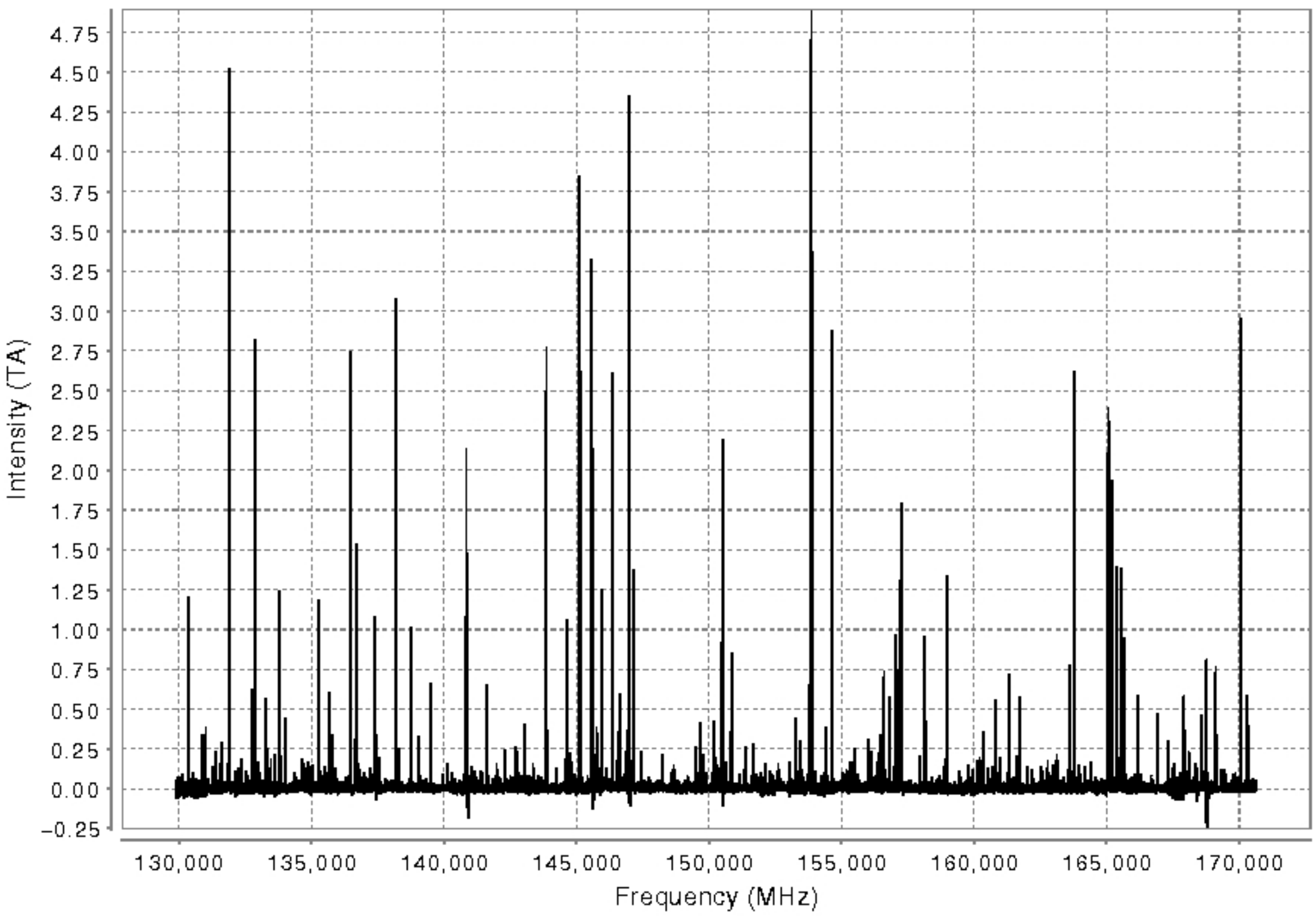